\documentclass[12pt]{article}
\usepackage{amsmath,amssymb,latexsym}
\pdfoutput=1

\usepackage[english]{babel}

\usepackage{graphicx}
\usepackage{color}

\usepackage{ifpdf}

\ifpdf
\usepackage[pdftex]{hyperref}
\else
\usepackage[hypertex,ps2pdf,dvips,colorlinks,urlcolor=blue,citecolor=blue]{hyperref}
\fi

\newcommand\be{\begin{equation}}
\newcommand\bea{\begin{align}}
\newcommand\bes{\begin{subequations}}
\newcommand\esu{\end{subequations}}
\newcommand\ee{\end{equation}}
\newcommand\eea{\end{align}}

\newcommand{\ud}          {\mathrm d}
\newcommand{\e}          {\mathrm e}

\newcommand\lam             {\lambda}

\renewcommand{\a}          {\alpha}

\newcommand\doi[2]        {\href{http://dx.doi.org/#1}{#2}}

\begin{document}

\begin{center}
{\Large\bf Stationary entanglement  entropies following an interaction quench in 1D Bose gas}
\\[2.1em]
\bigskip

{\large \bf Mario Collura, M\'arton Kormos, Pasquale Calabrese}
\null

\noindent 

{\it 

Dipartimento di Fisica dell'Universit\`a di Pisa and INFN, 56127 Pisa, Italy}

\date{\today}

\end{center}

\begin{abstract}

We analyze the entanglement properties of the asymptotic steady state after a quench from free to hard-core bosons in one dimension. The R\'enyi and von Neumann entanglement entropies are found to be extensive, and the latter coincides with the thermodynamic entropy of the Generalized Gibbs Ensemble (GGE). Computing the spectrum of the two-point function, we provide exact analytical results both for the leading extensive parts and the subleading terms for the entropies as well as for the cumulants of the particle number fluctuations. 
We also compare the extensive part of the entanglement entropy with the thermodynamic ones, showing that the GGE entropy 
equal the entanglement one and it is the double of the diagonal entropy.

\end{abstract}


\newpage

\section{Introduction}

Entanglement is a fundamental characteristic of quantum mechanics and it is the main feature distinguishing 
the quantum from the classical world. Over the last decade, it has become clear that quantum entanglement provides important information in many-body systems, for example in connection with criticality and 
topological order, see e.g. Ref. \cite{rev} for reviews.
For example, it has been understood that the amount of entanglement contained in a quantum system is the main limitation to the efficiency of numerical methods based on tensor network states \cite{swvc-08,cv-09}, such as the celebrated density matrix renormalization group. 
 
In pure quantum states, von Neumann and R\'enyi entanglement entropies of the reduced density
matrix $\hat\rho_A$ of a subsystem $A$ turned out to be very useful measures of entanglement.  
R\'enyi entanglement entropies are defined as 
\be
S^{(\a)}_A=\frac1{1-\a}\ln{\rm Tr}\,\hat\rho_{A}^\a\, .
\label{Sndef}
\ee 
For $\a\to1$ this definition gives the most commonly used von
Neumann entropy 
\be
S_A=-{\rm Tr}\,{\hat \rho_A\ln\hat \rho_A}\,,
\ee 
while for $\a\to\infty$
is the logarithm of the largest eigenvalue of $\hat\rho_A$, also called single copy entanglement \cite{sce}.
Furthermore, the knowledge of the $S^{(\a)}_A$ for different $\alpha$ characterizes the
full spectrum of non-zero eigenvalues of $\hat \rho_A$ \cite{cl-08}.
Just to quote two very important results, in the ground state of 1D systems whose continuum limit is conformally invariant,
the entanglement entropy grows logarithmically with subsystem size and the  pre-factor is proportional 
to the central charge \cite{Holzhey,cc-04,c-lec},
while in 2D systems displaying topological order (such as quantum Hall systems) 
a subleading term is the topological charge of the theory \cite{topo}.

Another field where entanglement entropy has been playing a crucial role is the out of equilibrium 
dynamics of quantum systems, such as after a quench of a Hamiltonian parameter in a closed system.
Indeed, based on results from conformal field theory \cite{cc-05,cc-07b,ds-11}
and on analytical \cite{cc-05,fc-08,ep-12} or numerical calculations \cite{dmcf-06,lk-08,ep-08,ep-08b,isl-09,va-12,cc-13,sled-13,hgf-09} 
for specific models, it is known that the entanglement entropy 
grows linearly with time for a  global quench, while at most logarithmically for a local one. 
As a consequence, a local quench can be effectively simulated with tensor networks up to large times, while for a global quench one can access only relatively short time dynamics.

Furthermore, the extensive behavior of the entanglement entropy for infinite time after a global quantum quench is reminiscent of a thermodynamic entropy. This connection is in fact rather natural: long time after a quench, the reduced density matrix of a subsystem is commonly accepted to be the one of the mixed state compatible with all the {\it local} integrals of motion, i.e. a generalized Gibbs ensemble (GGE) for an integrable 
system \cite{gg,mwn-07,cc-06,c-06,cdeo-08,bs-08,scc-09,fm-10,sfm-12,CEFII,eef-12,se-12,fe-13,p-13,fe-13b} 
and a thermal ensemble for a generic system \cite{kww-06,rdo-08,r-09,bkl-10,bch-11,gme-11,rs-12,sks-13}, 
see also Ref. \cite{revq} for a review.
In this construction, in order to establish the existence of a steady state, one first takes the thermodynamic limit (TDL) for the entire system and, only after this, one can consider a large subsystem \cite{CEFII}. 
Within this construction it is basically tautological that the thermodynamic entropy (i.e. the von Neumann entropy of the 
`final mixed state') must be equal to the entanglement one.
However, other entropies have been proposed in the literature \cite{p-11}, which can appear more natural for finite systems
(and are expected to coincide for non-integrable systems \cite{g-13}).
For these reasons it is worth investigating whether the infinite time entanglement entropy coincides with the 
thermodynamic entropy in some explicitly calculable cases. 
Up to now, this problem has been considered in great detail only for the 1D Ising model after a quench of the transverse 
field \cite{fc-08,g-13,CEF,f-13,rsm-10,gcg-11}, and only marginally in a few other cases \cite{g-13,spr-11,sr-10,v-e} 
(often in the equivalent formulation of the inverse participation ratio).

In this paper we study another exactly solvable instance, the quench from free to hard-core bosons in the continuum, for which an analytical solution was provided only recently by the present authors \cite{kcc-13,ksc-13} (see also \cite{nwbc-13}).
The computation of the entanglement entropy for arbitrary times is highly non-trivial because  
Wick's theorem does not apply \cite{kcc-13}, however for infinite time its validity is restored allowing for an exact calculation of the entropies as detailed in the following. 

The manuscript is organized as follows. In Section 2 we review some recent results on a special interaction quench in the Lieb--Liniger model, and we give a brief summary of the relevant formulae to compute entanglement entropies and particle fluctuations. We obtain our analytic results in two different ways. In Section 3 we present a direct method, while in Section 4 we introduce a novel approach based on the spectrum of the two-point function. After comparing the entanglement entropies with the thermodynamic entropy of the GGE in Section 5, we give our conclusions in Section 6.

\section{The model and the quantities of interest}

We consider the Lieb--Liniger model, a one-dimensional Bose gas with pairwise delta interaction
on a ring of circumference $L$ with periodic boundary conditions (PBC), i.e. with Hamiltonian \cite{LiebPR130}
\be\label{HLL}\hspace{-2mm}
H = \int_{0}^{L} \ud x \big[\partial_x \hat\phi^{\dagger} (x) \partial_x \hat\phi (x) 
+ c\, \hat\phi^{\dagger} (x) \hat\phi^{\dagger} (x) \hat\phi (x) \hat\phi(x) \big]\,,
\ee
where $\hat\phi(x)$ is a canonical boson field, $c$ the coupling constant and we set $\hbar=2m=1$.
 We are interested in the TDL, when $N,L\to\infty$ with the particle density $n=N/L$ fixed.
The Lieb--Liniger model is integrable for arbitrary value of the interaction parameter $c$, 
but, despite of the many approaches and results in the 
literature \cite{cro,fcc-09,grd-10,mc-12,ck-12,a-12,ce-13,bck-13,csc-13a,m-13,ds-13,ksc-13,nwbc-13}, the general non-equilibrium 
quench dynamics (e.g. a quench from arbitrary $c_0$ to arbitrary $c$) is still beyond reach.

For this reason, we consider the easiest quench dynamics in the model, which is the one from 
initial $c=0$ (free bosons) to final $c=\infty$ (impenetrable bosons).  
To be more specific, we prepare the many-body system in the $N$-particle ground state of the free boson Hamiltonian  
given by Eq. (\ref{HLL}) with $c=0$. 
At time $t=0$, we suddenly turn on an infinitely strong interaction, and the evolution 
is governed by the Hamiltonian (\ref{HLL}) with $c=\infty$.
The Jordan--Wigner transformation
\be
\hat{\Psi}(x)  =  \exp\left\{i\pi\int_{0}^{x}\ud z\hat{\Phi}^{\dag}(z) \hat{\Phi}(z)\right\}\, \hat{\Phi}(x)
\label{JWeq}
\ee
maps the hard-core boson Hamiltonian onto the free fermionic one \cite{TG}.
This dynamics has been studied numerically in Ref. \cite{grd-10} and analytically in Ref. \cite{kcc-13}. One of the main results of the latter is that for finite times Wick's theorem does not 
apply and each multi-point correlator should be calculated separately. However, for infinite time, 
since the stationary state is described by the density matrix
$\rho_{GGE} = Z^{-1}\exp[-\int dk \lambda(k)\hat{n}(k)/2\pi]$ which is diagonal in momentum modes $\hat{n}(k)$, 
Wick's theorem is restored and all multi-point correlators can be determined in terms of the fermionic
two-point function \cite{kcc-13,ksc-13}
\be\label{GGE_corr}
C(x-y)\equiv \langle \Psi^\dagger (x,t=\infty) \Psi(y,t=\infty) \rangle
=C_{\text{GGE}}(x-y) = n \mathrm{e}^{-2n|x-y|}\,,
\ee 
where $n$ is the particle density. 

It is important to stress that since the Jordan-Wigner mapping (\ref{JWeq}) guarantees that 
the fermions in a given interval are functions only of the bosons in the same interval,
the entanglement entropies of a single interval for the impenetrable bosons and for free fermions 
{\it do coincide}.  
This is not true anymore in the case of more disjoint intervals because of the presence of a bosonization string in Eq. (\ref{JWeq}), 
analogously to what happens in spin-chains \cite{twoint}.

\subsection{Entanglement entropies and particle fluctuations}

For a one-dimensional quantum gas that can be mapped to a non-interacting fermion system,
Wick's theorem allows for an exact representation of the bipartite entanglement entropies of any spatial subsystem in terms of the two-point fermion correlator \cite{cmv-11,cmv-11b}. 
Let us denote this two-point function by $C(z,z')\equiv\langle\hat{\Psi}^{\dag}(z)\hat{\Psi}(z')\rangle.$ 
The reduced density matrix of a spatial subsystem $A$ is 
\be
\hat\rho_A \propto \exp \Big(- \int_A \ud y_1 \ud y_2 \Psi^\dagger(y_1) {\cal H}(y_1,y_2) \Psi(y_2)\Big)\,,
\label{rhoa}
\ee
where ${\cal H}=\ln [(1-C)/C]$ and the normalization constant is fixed requiring ${\rm Tr}\hat\rho_A=1$.
This equation can be straightforwardly seen as the continuum limit of the formula for lattice free fermions \cite{p-lat,ep-rev}, but 
has also been  obtained directly in the continuum path integral formalism \cite{p-lat}. 
At this point the integer powers, and hence the R\'enyi entropies, of this reduced density matrix are given by Wick's theorem as 
\be
S^{(\alpha)}_{A} =\frac{1}{1-\alpha} \mathrm{Tr}\,\ln\left[\mathbb{C}_A^\alpha+(1-\mathbb{C}_A)^\alpha\right]\,,
\label{SvsC}
\ee
where $\mathbb{C}_A$ is the restriction of the fermionic correlation to the subsystem $A$.

In order to be more explicit, the trace of the powers of the restricted correlation function is defined as
\be\label{C_traces}
\mathrm{Tr}\,\mathbb{C}^k_{A}\equiv \int_A \ud z_1\dots \ud z_k C(z_1,z_2) C(z_2,z_3) \dots C(z_k,z_1)\,.
\ee
Introducing the matrix 
\be
\mathbb{E}_A=\mathbb{C}_A(1-\mathbb{C}_A)\,,
\label{Edef}
\ee
the trace log in Eq. (\ref{SvsC}) can be recast in the form \cite{cmv-12}
\be
-\mathrm{Tr}\,\ln\left[\mathbb{C}^\alpha_A+(1-\mathbb{C}_A)^\alpha\right]=\\
\sum_{k=1}^\infty \frac{4^k}k \mathrm{Tr}\,\mathbb{E}^k_A \sum_{p=1}^{\lfloor\alpha/2\rfloor}\cos^{2k}\left(\frac{2p-1}{2\alpha}\pi\right)\,.
\label{SvsE}
\ee
The previous formulas permit us to write all R\'enyi entropies of integer order in terms of the integer powers 
$\mathrm{Tr}\,\mathbb{C}^k_{A}$ in Eq. (\ref{C_traces}).
When a closed analytic form for all integer $\alpha$ has been found, one can use the replica trick and search for an 
analytic continuation to non-integer $\alpha$, whose limit for $\alpha\to1$ would give the desired von Neumann 
entanglement entropy.
This is the first approach we will use in the following to determine the entanglement entropy. 

An alternative way to obtain the entanglement entropies directly for any real $\alpha$ is provided by finding the spectrum of the reduced correlation matrix.  
Indeed, if one knows all the eigenvalues $\lambda_m$ of $\mathbb{C}_{A}$ (which we assume 
to be discrete for simplicity, as will be in the case of our interest), Eq. (\ref{SvsC}) can be simply rewritten as
\be\label{entropies}
S^{(\alpha)}_A
= \sum_{m}e_{\alpha}(\lambda_{m}), \qquad 
e_{\alpha}(\lambda)\equiv\frac1{1-\alpha}\ln[\lambda^\alpha+(1-\lambda)^\alpha].
\ee
For $\alpha=1$ this formula gives the von Neumann entropy 
\be
S_A= -\sum_{m} [ \lambda_m \ln \lambda_m + (1-\lambda_m) \ln(1-\lambda_m)]\,,
\ee
and for $\alpha\to\infty$ the single copy entanglement
\be
S^{(\infty)}_A = -\sum_{m}  \ln \left( \frac{1}{2} + \left| \lambda_m - \frac{1}{2} \right| \right)\,.
\label{scedef}
\ee

Usually \cite{cmv-11b,jk-04} the spectrum of reduced correlation matrix is calculated by introducing  
the Fredholm determinant
\be
\mathcal{D}_{A}(\lambda) = \det [\lambda \delta_A(z-z') - {\mathbb C}_A(z,z')]\,,
\ee
where also the identity $\delta(z-z')$ is restricted to the subsystem $A$.
If one is able to calculate (the asymptotic behaviour of) $\mathcal{D}_{A}(\lambda)$, 
the entanglement entropies are given by Eq.\ \eqref{SvsC} as the integral \cite{cmv-11b,jk-04}
\be\label{entropies_contour}
S^{(\alpha)}_{A}
= \oint\frac{\ud\lambda}{2\pi i}e_{\alpha}(\lambda)\frac{\ud}{\ud\lambda}\ln \mathcal{D}_{A}(\lambda),
\ee
over a contour which encircles the segment $[0,1]$.
However, we will not exploit this method in the following because in the 
present case we found it easier to directly diagonalize the reduced correlation matrix. 


\subsection{Particle fluctuations}
The R\'enyi entropies characterize the non-trivial connections between different parts of an extended quantum system. 
For systems which can be mapped to free fermions as the present one, 
the entanglement entropies can be related to the even cumulant $V^{(2k)}_{A}$ 
of the particle-number distribution \cite{kl-09,song1,song2,cmv-12cum}
\be\label{cumulant_0}
V^{(k)}_{A} = (-i\partial_{\lambda})^{k}\ln\langle \mathrm{e}^{i\lambda \hat{N}_{A}} \rangle |_{\lambda = 0}\,,
\ee
where
\be
\hat{N}_{A} = \int_{A}\! \ud z \, \hat{\Psi}^{\dag}(z)\hat{\Psi}(z)\,,
\ee
is the operator counting the number of particles in the interval $A=[x,y]$. 
Indeed, it has been shown that the following formal expansion holds \cite{song2}
\be
\label{Salpha_cumulant}
S^{(\alpha)}_{A}   =  \sum_{k=1}^{\infty} s^{(\alpha)}_{k} V^{(2k)}_{A},\qquad 
s^{(\alpha)}_{k}   =   \frac{(-1)^{k}(2\pi)^{2k}2\zeta[-2k,(1+\alpha)/2]}{(\alpha-1)\alpha^{2k} (2k)!}\,,
\ee
where $\zeta[n,x]\equiv \sum_{k=0}^{\infty}(k+x)^{-n}$ is the generalized Riemann zeta function. 
In particular, exploiting the definition of the cumulant in Eq. (\ref{cumulant_0}), one has
\begin{align}
V^{(k)}_{A} & =  (-i\partial_{\lambda})^{k} G(\lambda,\mathbb{C}_{A})|_{\lambda=0}\,, \nonumber\\
G(\lambda,\mathbb{C}_{A}) & =  \mathrm{Tr} \ln [ \mathbb{I} - (1-\mathrm{e}^{i\lambda}) \mathbb{C}_{A} ]\,.
\end{align}
Also in this case we can use the spectrum $\{\lambda_{m}\}$ of the correlation function obtaining
\be
V^{(k)}_{A} =  \sum_{m} \left. (-i\partial_{\lambda})^{k}  \ln [ 1 - (1-\mathrm{e}^{i\lambda}) \lambda_{m} ] \right|_{\lambda=0}\,,
\ee
which can be rewritten in terms of polylogarithm functions $\mathrm{Li}_{k}(z)=\sum_{n=1}^{\infty}z^n / n^k$ as \cite{csc-13}
\be\label{cumulant_spectr}
V^{(k)}_{A} = -\sum_{m}\mathrm{Li}_{1-k}\left(\frac{\lambda_m}{\lambda_m-1}\right)=
- {\rm Tr}\, \mathrm{Li}_{1-k}\left(\frac{{\mathbb C}_A}{{\mathbb C}_A-1}\right)\,. 
\ee

\section{Direct approach}

In this section we consider the entanglement entropies and particle fluctuations of a subsystem $A$ consisting 
of an interval of length $\ell$ in the infinite system employing a brute force direct computation of the 
traces of the restricted correlation matrix, as in Eq.\ \eqref{C_traces}.
For low powers, making use of the  two-point function in Eq. (\ref{GGE_corr}), the integrals in Eq. (\ref{C_traces}) 
can be easily calculated and for the first five we obtain
\begin{subequations}
\label{trC}
\begin{align}
\mathrm{Tr}\,\mathbb{C}_{A} & =  n\ell \,,\\
\mathrm{Tr}\,\mathbb{C}^2_{A} & =  (4n\ell-1+\mathrm{e}^{-4n\ell})/8\,, \\
\mathrm{Tr}\,\mathbb{C}^3_{A} & =  [6n\ell-3+\mathrm{e}^{-4n\ell}(6n\ell+3)]/16\,,\\
\mathrm{Tr}\,\mathbb{C}^4_{A} & = [40n\ell-29+4\mathrm{e}^{-4n\ell}(16n^2\ell^2+20n\ell+7)+\mathrm{e}^{-8n\ell}]/128\,,\\
\mathrm{Tr}\,\mathbb{C}^5_{A}  & = 5[ ( 42n\ell - 39 ) + 4 \mathrm{e}^{-4n\ell} ( 4n\ell + 3 )(4n^2\ell^2+6n\ell+3) +\mathrm{e}^{-8n\ell} (6n\ell+3 ) ]/24\,,
\end{align}
\end{subequations}
where we recall that $\ell$ is the length of the interval $A.$

Based on the first few powers we are led to the conjecture 
\begin{align}
\mathrm{Tr}\,\mathbb{C}^k_{A} &= \frac1{2^{2k-2}}\binom{2k-2}{k-1} \,n\ell+
\frac1{2^{2k-1}}\binom{2k-1}{k-1}-\frac12+\mathcal{O}(e^{-4n\ell}) \nonumber \\
&=
\frac{\Gamma(k-\frac12)}{\sqrt{\pi}\,\Gamma(k)}n\ell+
\frac{\Gamma(k+\frac12)}{\sqrt{\pi}\,\Gamma(k+1)}-\frac12+\mathcal{O}(e^{-4n\ell})\,,
\label{conjCk}
\end{align}
which we prove in the next section using results on the spectrum of $\mathbb{C}_{A}$.


The conjecture (\ref{conjCk}) contains all ingredients to calculate the leading and subleading contributions to the 
R\'enyi entanglement entropy of arbitrary order. 
Indeed, Eq.\ (\ref{conjCk}) implies that for the matrix ${\mathbb E}_A$ in Eq.\ (\ref{Edef}) we have
\be
\mathrm{Tr}\,\mathbb{E}_A^k = 
\frac{\Gamma(k-\frac12)}{2^{2k-1}\sqrt{\pi}\,\Gamma(k)}\left(n\ell+\frac12-\frac1{4k}\right) +\mathcal{O}(e^{-4n\ell})
\equiv  e^1_k n\ell +e^0_k+O(e^{-4n\ell})\,.
\label{Eexp}
\ee

By straightforward summation of Eq. (\ref{SvsE}), $\mathrm{Tr}\,\mathbb{E}_A^k$ gives all the R\'enyi entropies of 
low integer order, as for example
\begin{align}
S^{(2)}_A&=(4-2\sqrt{2})n\ell +\ln(24-16\sqrt{2})+\mathcal{O}(e^{-4n\ell})=
(4-2\sqrt{2})n\ell +2\ln\left[8\sin\left(\frac\pi8\right)\right]+\mathcal{O}(e^{-4n\ell}) ,\nonumber\\
S^{(3)}_A&=n\ell +\ln(4/3) +\mathcal{O}(e^{-4n\ell})\,,\nonumber\\
3S^{(4)}_A&=(8-2\sqrt{2(2+\sqrt{2})})n\ell +\ln(256) - 2 \ln(4 + \sqrt{2} + 2 \sqrt{2 (2 + \sqrt{2} ) }) +\mathcal{O}(e^{-4n\ell})\nonumber\\
&=\left(8-\frac2{\sin(\pi/8)}\right)n\ell+\ln2 + 2 \ln\left[8\tan\left(\frac{\pi}{16}\right) \tan\left(\frac{3 \pi}{16}\right)\right] +\mathcal{O}(e^{-4n\ell})\,.
\end{align}

In order to be systematic and give close formulas for arbitrary integer and real $\alpha$, let us expand
the R\'enyi entropies in powers of $\ell$ as
\begin{align}
S^{(\alpha)}_A&= s^1_\alpha n\ell + s^0_\alpha +{\cal O}(e^{-4n\ell})\,.
\label{Saexpl}
\end{align}
The series coefficients $s^a_\alpha$ are clearly related to the factors $e^{a}_k$ in Eq. (\ref{Eexp}).
According to Eq.\ (\ref{SvsE}) this relation reads 
\be
s^a_\alpha=\frac1{\alpha-1}
\sum_{k=1}^\infty \frac{4^k}k  e^a_k \sum_{p=1}^{\lfloor\alpha/2\rfloor}\cos^{2k}\left(\frac{2p-1}{2\alpha}\pi\right)\,.
\label{sa11}
\ee
A simpler analytic expression can be obtained by exchanging the order of the two summations.
Let us first consider the term linear in $n\ell$, in which we can use the formula for the sum over $k$ 
\be
\sum_{k=1}^\infty e_k^1\frac{4^k}{k}\cos^{2k}x=
\sum_{k=1}^\infty \frac{\Gamma(k-\frac12)}{2^{2k-1}\sqrt{\pi}\,\Gamma(k)}\frac{4^k}{k}\cos^{2k} x=4(1-\sin x)\,,
\ee
which inserted in Eq. (\ref{sa11}) provides
\be
s^1_\alpha=\frac4{\alpha-1} \sum_{p=1}^{\lfloor\alpha/2\rfloor}  \left(1 + \sin \frac{\pi(1 - 2 p )}{2 \alpha}\right)\\
=2\frac{\csc \frac\pi{2\alpha}-\alpha}{1-\alpha}\,,
\ee
giving, in particular,  $s^1_1=2$.

The calculation of the subleading term (order one  in $n\ell$) is more complicated. 
For $ x\in (0,\pi/2]$, we have 
\be
\sum_{k=1}^\infty e_k^0\frac{4^k}{k}\cos^{2k}x=
\sum_{k=1}^\infty \frac{\Gamma(k-\frac12)}{\sqrt{\pi}\,\Gamma(k+1)} \left(1-\frac1{2k}\right)\cos^{2k} x=
-4\ln \sin \Big(\frac{x}2+\frac\pi4\Big)\,,
\ee
and so 
\be
s^0_\alpha=\frac4{1-\alpha} \sum_{p=1}^{\lfloor\alpha/2\rfloor}\ln \sin\left(\pi\frac{\alpha-1+2p}{4\alpha}\right)
= \frac4{1-\alpha} \sum_{p=1}^{\lfloor\alpha/2\rfloor}\ln \cos\left(\pi\frac{1+\alpha-2p}{4\alpha}\right)\,.
\ee
In order to perform explicitly the above sum, we use the integral \cite{GR}
\be
\ln \cos \frac{\pi a}b=-2 \int_0^\infty \frac{\ud x}x \frac{\sinh^2(ax)}{\sinh(bx)}\,,\qquad b> 2|a|,
\ee
and obtain
\be
s^0_\alpha=-\frac8{1-\alpha} \sum_{p=1}^{\lfloor\alpha/2\rfloor}\int_0^\infty \frac{\ud x}x \frac{\sinh^2[(1+\alpha-2p)x]}{\sinh(4\alpha x)}\,.
\ee
Exchanging the order of sum and integral, the sum can be performed as
\be
-4 \sum_{p=1}^{\lfloor\alpha/2\rfloor}\sinh^2((1+\alpha-2p)x)=\alpha -{\rm csch}(2 x) \sinh(2 \alpha x)\,,
\ee
so that
\be
s^0_\alpha=\frac2{1-\alpha}\int_0^\infty \frac{\ud x}x \frac{\alpha -{\rm csch}( x) \sinh( \alpha x)}{\sinh(2\alpha x)}\,.
\label{sarep}
\ee
Now the limit $\alpha\to1$ can be taken straightforwardly obtaining
\be
s^0_1=2\int_0^\infty \frac{\ud x}x \frac{ x \coth  x-1}{\sinh (2 x)}= 2\ln2-1\,.
\ee

Collecting the linear and the subleading terms, we arrive to the main result of the section, i.e. 
the analytic expression for the R\'enyi entropies
\be
S^{(\alpha)}_A= 2\frac{\csc \frac\pi{2\alpha}-\alpha}{1-\alpha} n\ell +
 \frac2{1-\alpha}\int_0^\infty \frac{\ud x}x \frac{\alpha -{\rm csch}( x) \sinh( \alpha x)}{\sinh(2\alpha x)} +{\cal O}(e^{-4n\ell}),
\label{SAafin}
\ee
which in the limit $\a\to1$ gives the von Neumann entanglement entropy 
\be
S_A= 2 n\ell +2\ln2-1+\mathcal{O}(e^{-4n\ell})\,.
\label{S11}
\ee


\begin{figure}[t]
\includegraphics[width=0.48\textwidth]{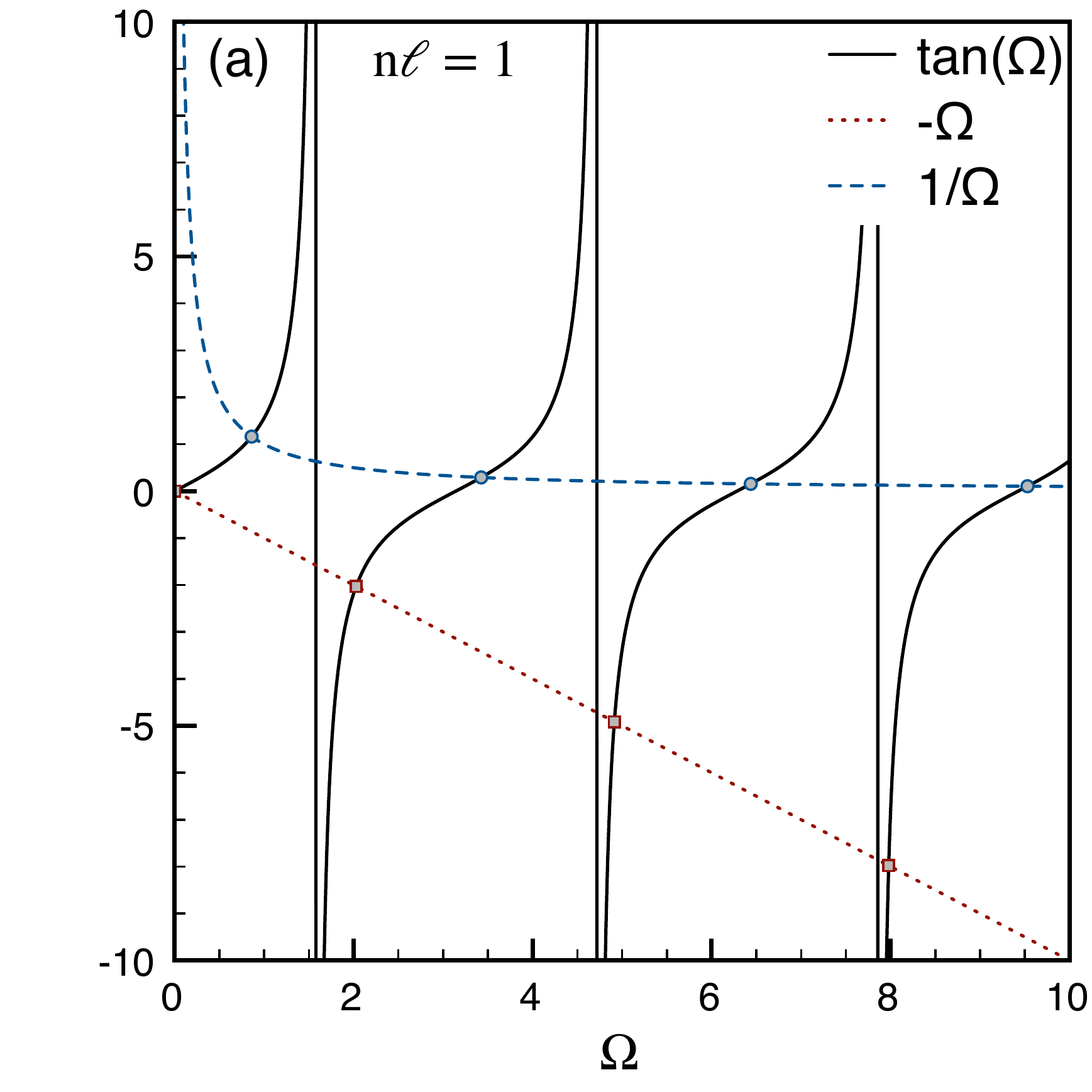} \includegraphics[width=0.48\textwidth]{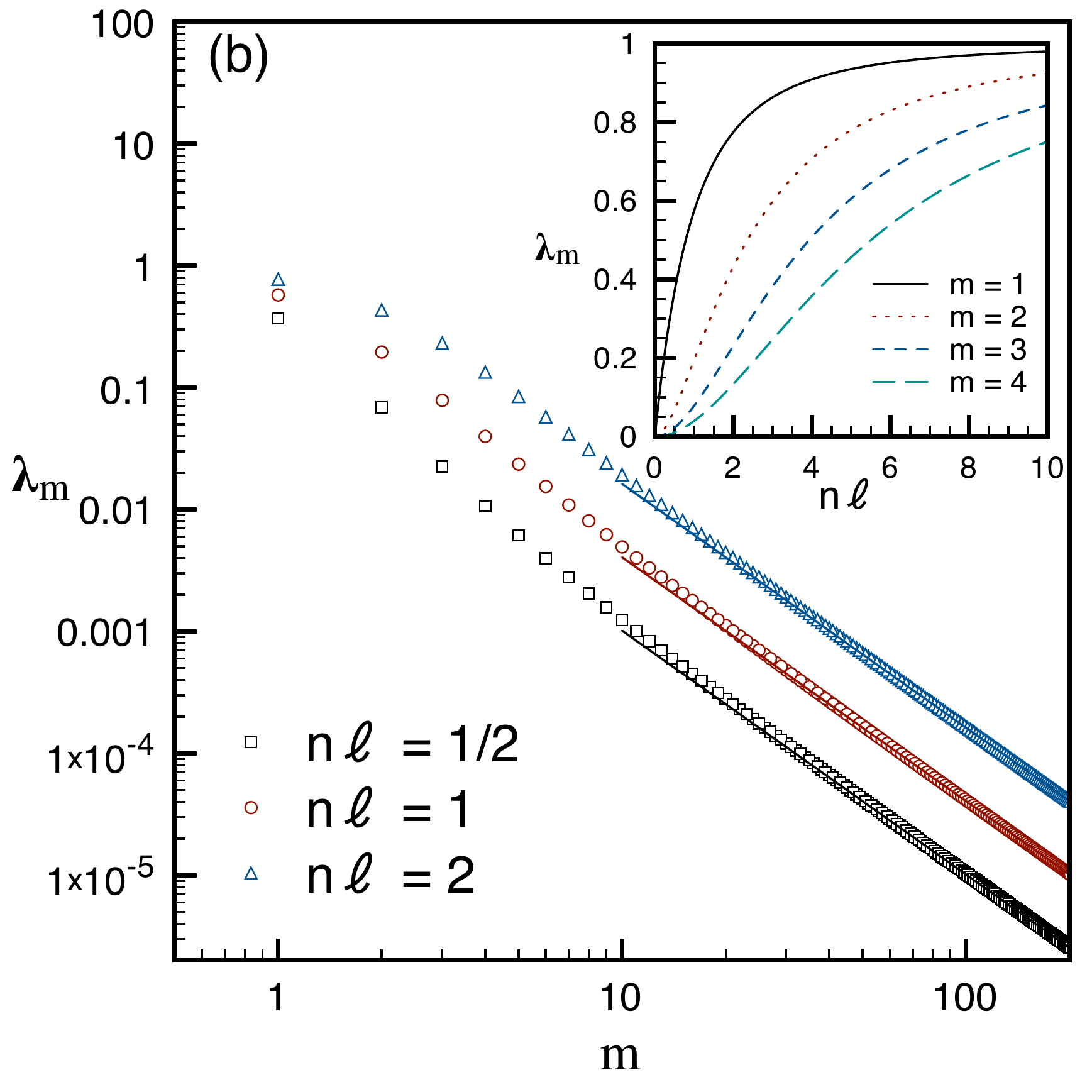}
\caption{(a) Pictorial representation of the first few solutions of Eq.\ \eqref{equation}. 
The trivial solution $\Omega=0$ is a spurious root which does not correspond to an eigenvalue of the original problem.
(b) The eigenvalues $\lambda_m$ vs.\ $m$ for different values of $n\ell$. The full lines are the asymptotic behavior in Eq.\ (\ref{lambda_asympt}). In the inset we report the behavior of the first four non-trivial eigenvalues as a function of the rescaled interval length $n\ell$. All the eigenvalues flow to $1$ for $n\ell\to\infty$.}
\label{fig1}
\end{figure}

\section{Spectrum of the correlation function}

An alternative way to compute the entanglement entropies and the particle fluctuations is based on the knowledge of the full spectrum of the restricted two-point fermionic correlation function ${\mathbb C}_A$. This allows for the numerically exact computation of the entanglement entropies for arbitrary subsystem size, and leads to exact integral formulas for the leading and subleading terms in $n\ell$ which agree with the results of the previous section.

The spectrum of the correlation in Eq.\ (\ref{GGE_corr}) restricted to an interval of length $\ell$ 
is given by  the continuum eigenvalue problem
\be\label{eigen_int}
\int_{0}^{\ell} \ud y\, n\, \mathrm{e}^{-2n|x-y|} v_{m} (y) = \lambda_{m} v_{m}(x)\,,
\ee
where, thanks to translational invariance, the integral depends only on the length $\ell>0$ of the integration interval. 
Notice that, since the exponential kernel is real and symmetric with norm smaller than one, 
the eigenvalues $\lambda_{m}$ are real and fall in the interval $[0,1]$.

As done in Ref. \cite{ep-13} for a different kernel,  taking the first two derivatives with respect to $x$ of the integral equation 
(\ref{eigen_int}), 
one can recast it as the following second order differential equation
\be\label{eigen_diff}
\partial^{2}_{x} v_{m}(x) = - \omega^{2}_{m} v_{m}(x)\,,
\ee
with $\omega^{2}_{m}\equiv 4n^2 (1/\lambda_{m}-1) \in \mathbb{R}$. 
The solutions of Eq.\ (\ref{eigen_diff}) are  
\be\label{eigenvectors_1}
v_{m}(x) = A_{m} \cos \omega_{m} x + B_{m} \sin \omega_{m} x\,.
\ee
The coefficients $A_{m}$ and $B_{m}$ as well as the `frequencies' $\omega_{m}$ are determined 
by the boundary conditions at $x=0$ and $x=\ell$ that are imposed by the integral equation. In particular, one has
\be
\partial_{x} v_{m} (0)  =  2n\,v_{m}(0)\,,\quad \partial_{x} v_{m} (\ell)  =  -2n\,v_{m}(\ell)\,,
\ee
which leads to the linear system
\be
\mathbf{M} \left( \begin{array}{c} A_m\\ B_m \end{array} \right) = 0\,,
\ee
with
\be
\mathbf{M}\equiv \left(\begin{array}{cc} 2n & -\omega \\ 
2n\cos \omega\ell-\omega\sin \omega \ell  & 2n\sin\omega\ell + \omega\cos\omega\ell \end{array}\right)\,.
\ee
In order to have a set of eigenfunctions building up the full eigensubspace, we have to impose the condition 
$\det \mathbf{M}=0$ which leads to the following equation for the rescaled frequencies 
$\Omega_{m}\equiv \omega_{m}/2n$, 
\be
\tan(2n\ell\, \Omega) = \frac{2\,\Omega}{\Omega^2 - 1}\,.
\label{equation}
\ee 
Using the trigonometric identity for $\tan(2x)$, 
all the solutions of Eq. \eqref{equation} can be rewritten as the union of the solutions of the following 
two independent equations
\be\label{eigenvalues_eqs}
\tan(n\ell\, \Omega) = - \Omega\,,\quad \tan(n\ell\, \Omega) = \frac{1}{\Omega}\,.
\ee
Finally, the eigenvalues of the integral equations are related to $\Omega_{m}$ by
\be
\lambda_{m} = \frac{1}{1+\Omega^{2}_{m}}\,.
\label{speclam}
\ee
The asymptotic behavior for large $m$ is easily obtained from Eq. (\ref{eigenvalues_eqs}):
\be\label{lambda_asympt}
\lambda_{m} \sim \Omega_{m}^{-2} \sim \left(\frac{\pi m}{2 n \ell}\right)^{-2},\quad m\gg1.
\ee
In the left panel of Fig.\ \ref{fig1} we give a pictorial representations of the first solutions of the Eq.\ (\ref{eigenvalues_eqs}), 
and in the right panel we report the numerically obtained eigenvalues $\lambda_{m}$ as function of $n\ell$.

\begin{figure}[t]
\includegraphics[width=0.48\textwidth]{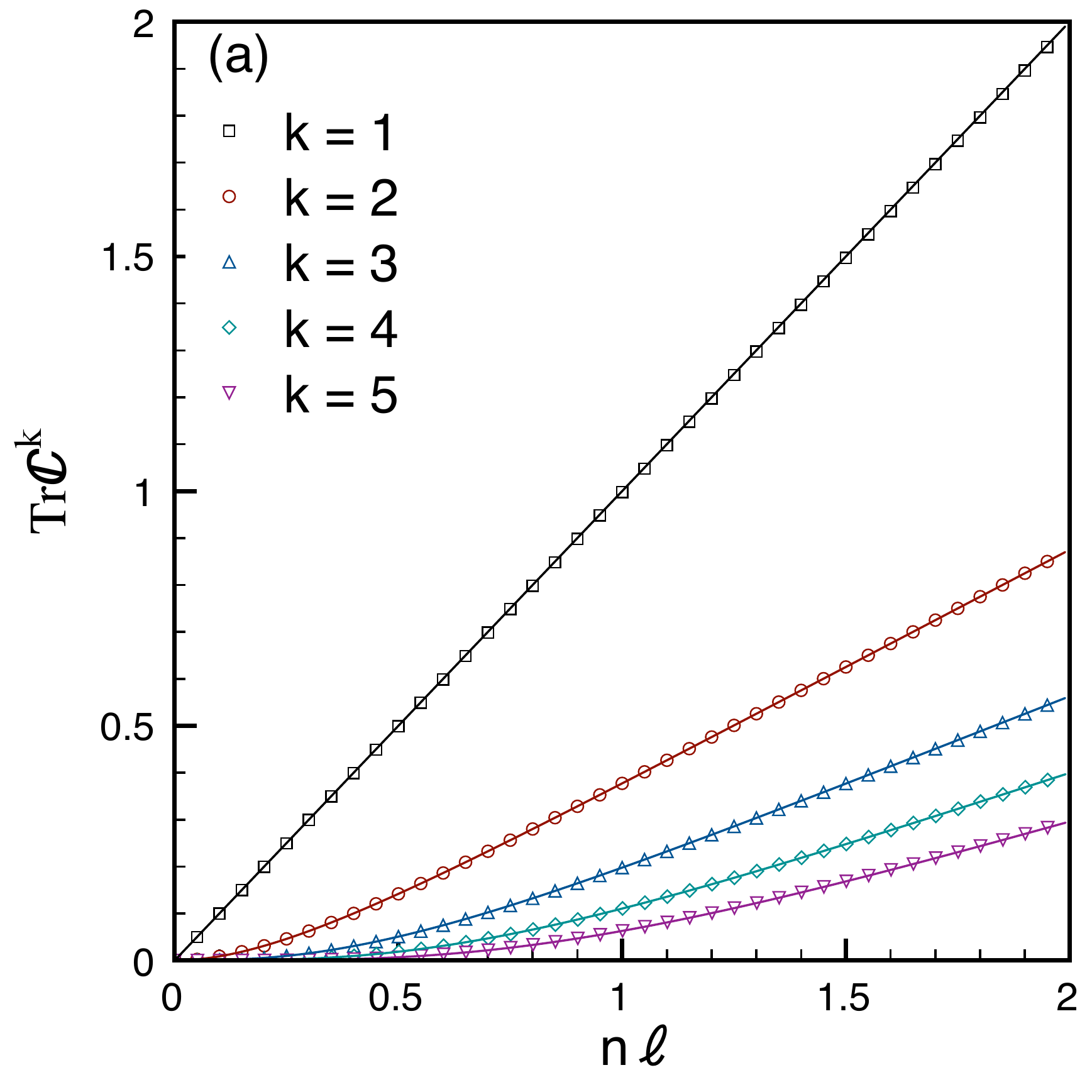} \includegraphics[width=0.48\textwidth]{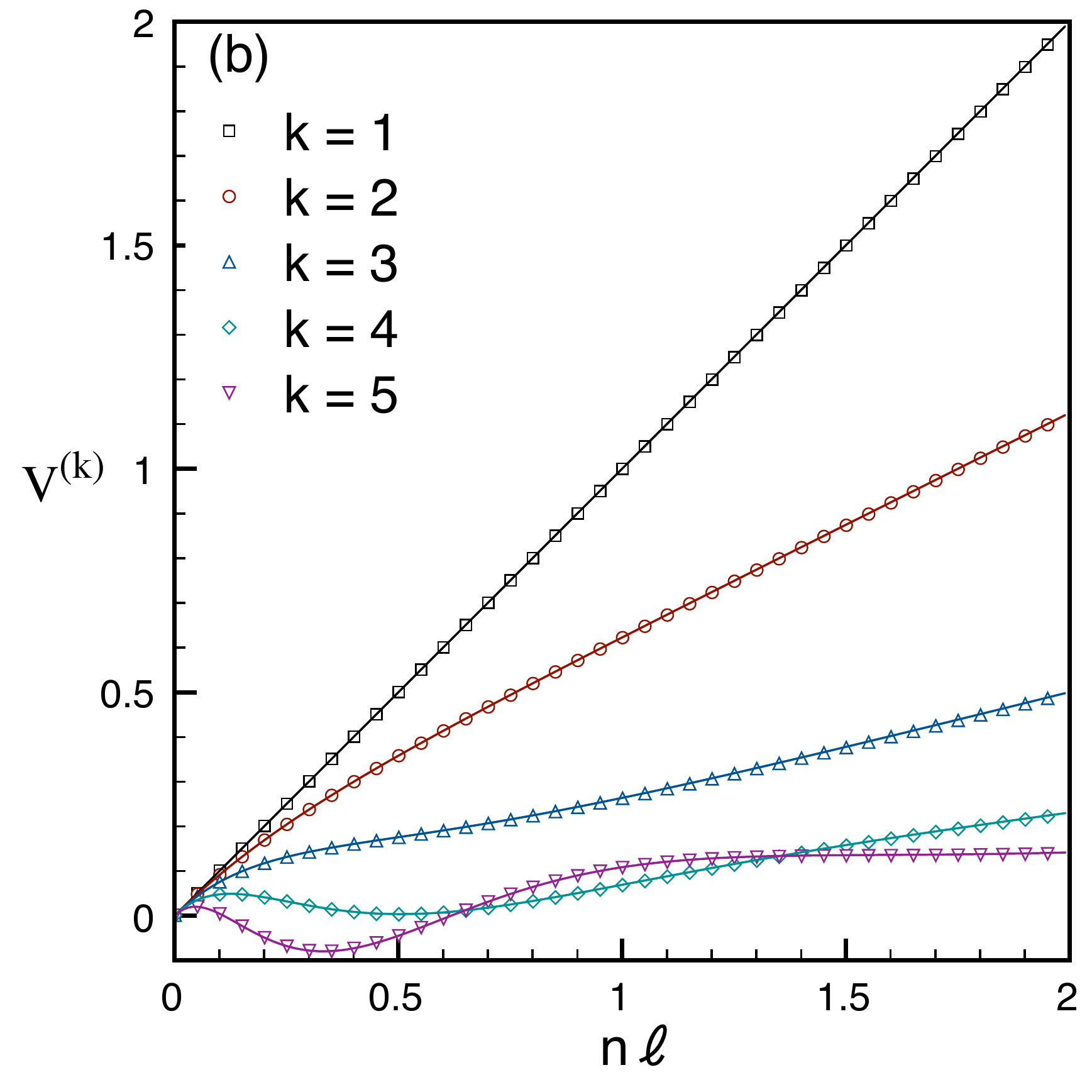}
\caption{
(a) The trace of the first five powers of the reduced correlation matrix as function of $n\ell$. 
The solid lines are the analytical formulae in Eq.\ \eqref{trC}, while the symbols represent the numerical evaluation 
using the spectrum $\{\lambda_{m}\}$ as in Eq. (\ref{speclam}). 
(b) The first five cumulants $V^{(k)}_{A}$; the lines and symbols are the same as in panel (a) comparing numerical and 
analytic results.
}
\label{fig2}
\end{figure}

 \subsection{Numerical results}

The eigenvalues $\lambda_m$ are easily found for arbitrary $n\ell$ by solving the two equations in (\ref{eigenvalues_eqs}),
as pictorially depicted in Fig.\ \ref{fig1}. 
In order to check the correctness of the solution for the spectrum of the correlation function, 
we first compute the traces ${\rm Tr}\, {\mathbb C}_A^k$ for the lowest values of $k$ obtained 
summing over the eigenvalues $\lambda_m$. 
In Fig.\ \ref{fig2} (left panel), these are compared with the analytical expressions in Eqs.\ (\ref{trC})  showing a 
perfect agreement for all values of $n\ell$.

\begin{figure}[t!]
\includegraphics[width=0.68\textwidth]{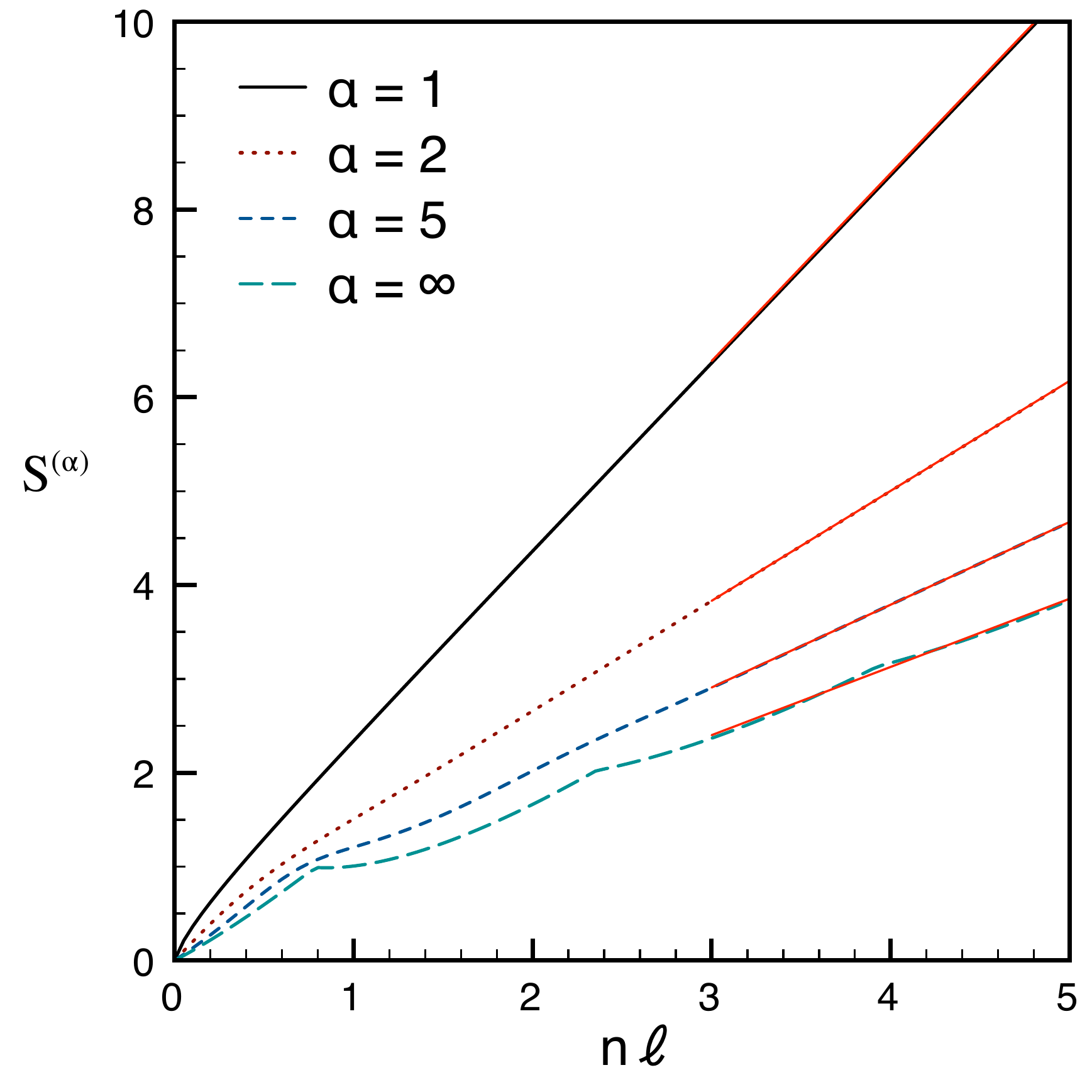}\\
\caption{Exact numerical evaluation of the R\'enyi entanglement entropy for $\alpha=1,2,5,\infty$ as function of 
$n\ell$. For large $n\ell$ these are perfectly described by the extensive asymptotic results reported as full (red) lines.
Notice the presence of some non-analytic points for $\alpha=\infty$.
\label{fig35}}
\end{figure}

In the same way, in Fig. \ref{fig2} (right panel), we compare the cumulants numerically evaluated via Eq.\ (\ref{cumulant_spectr}) with the following analytic form for the first five cumulants
\begin{subequations}
\begin{align}
V^{(1)}_{A} & =  \mathrm{Tr}\,\mathbb{C}_{A} \,,\\
V^{(2)}_{A} & =  \mathrm{Tr}\,\mathbb{C}_{A} - \mathrm{Tr}\,\mathbb{C}^{2}_{A}\,,\\
V^{(3)}_{A} & =  \mathrm{Tr}\,\mathbb{C}_{A} - 3\mathrm{Tr}\,\mathbb{C}^{2}_{A}
+2\mathrm{Tr}\,\mathbb{C}^{3}_{A}\,,\\
V^{(4)}_{A} & =  \mathrm{Tr}\,\mathbb{C}_{A} - 7\mathrm{Tr}\,\mathbb{C}^{2}_{A}
+12\mathrm{Tr}\,\mathbb{C}^{3}_{A}-6\mathrm{Tr}\,\mathbb{C}^{4}_{A}\,,\\
V^{(5)}_{A} & =  \mathrm{Tr}\,\mathbb{C}_{A} - 15\mathrm{Tr}\,\mathbb{C}^{2}_{A}
+50\mathrm{Tr}\,\mathbb{C}^{3}_{A}-60\mathrm{Tr}\,\mathbb{C}^{4}_{A}+24\mathrm{Tr}\,\mathbb{C}^{5}_{A}\,.
\end{align}
\end{subequations}

In Fig. \ref{fig35} we report the numerically evaluated entanglement entropies 
$S^{(1)}_{A}$, $S^{(2)}_{A}$, $S^{(5)}_{A}$, and $S^{(\infty)}_{A}$ for a subsystem of length $\ell$.
It is evident that for large enough $n\ell$, i.e. for $n \ell\geq 3$, the numerical results are very well described 
by the asymptotic formula with the linear (in $\ell$) and the constant term reported in Eq. (\ref{Saexpl}).
The asymptotic formula agrees quite well with the data also for relatively low values of $n\ell$ 
because, as in Eq. (\ref{Saexpl}), the corrections to this are exponentially small in $n\ell$.
This is very different from what is usually found in the ground state where power law corrections 
give sizeable effects also for much larger values of $\ell$.
It is also worth noticing that for $\alpha=\infty$, there is an infinite sequence of non-analytic points in $n\ell$ 
which reflect the presence of an absolute value in the sum of the eigenvalues $\lambda_m$ in Eq. (\ref{scedef}).
For finite $\alpha$, these singularities are smoothed out but some signs of their appearance are clear (see e.g. the curve for $\alpha=5$ in Fig. \ref{fig35}).

\begin{figure}[t!]
\includegraphics[width=0.48\textwidth]{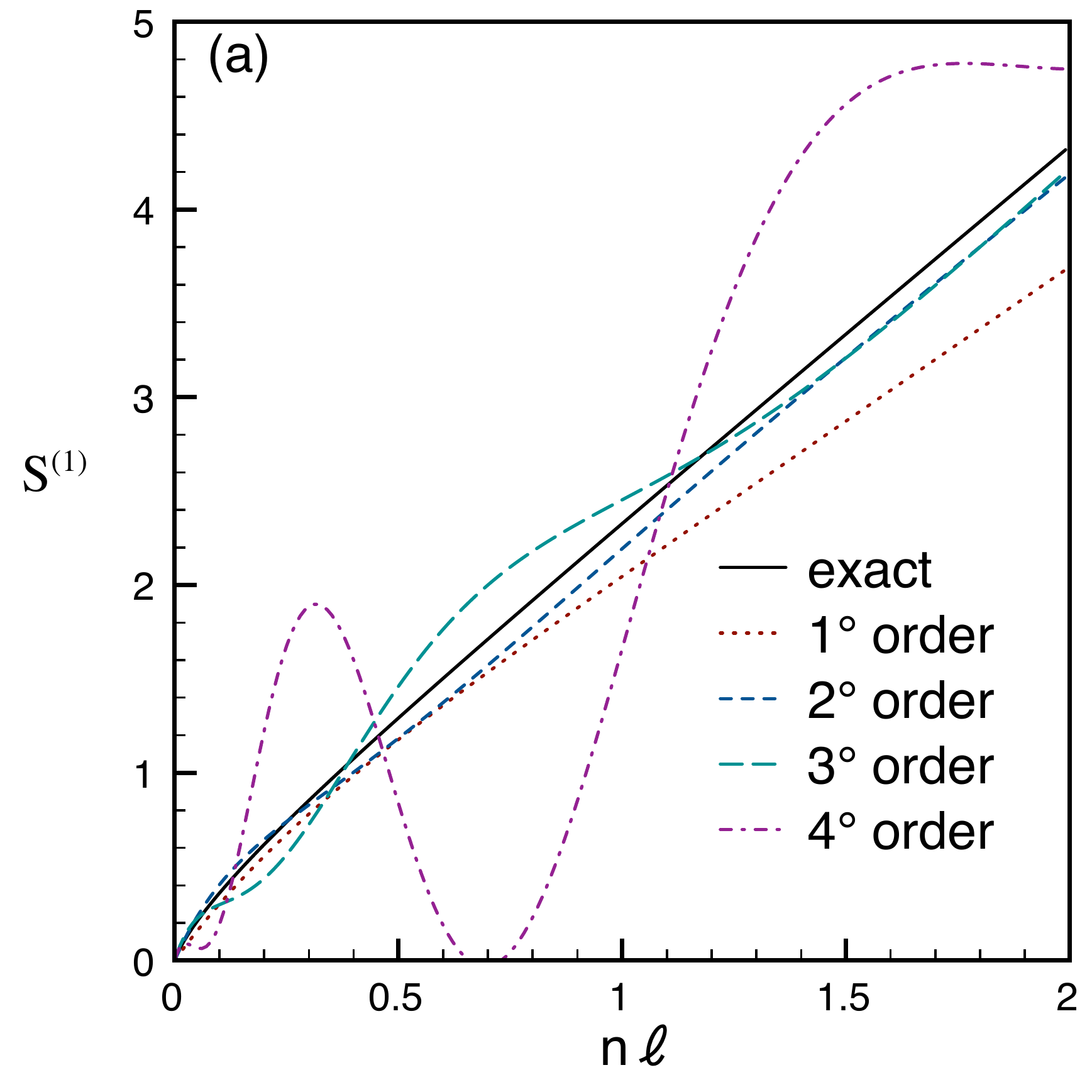}\includegraphics[width=0.48\textwidth]{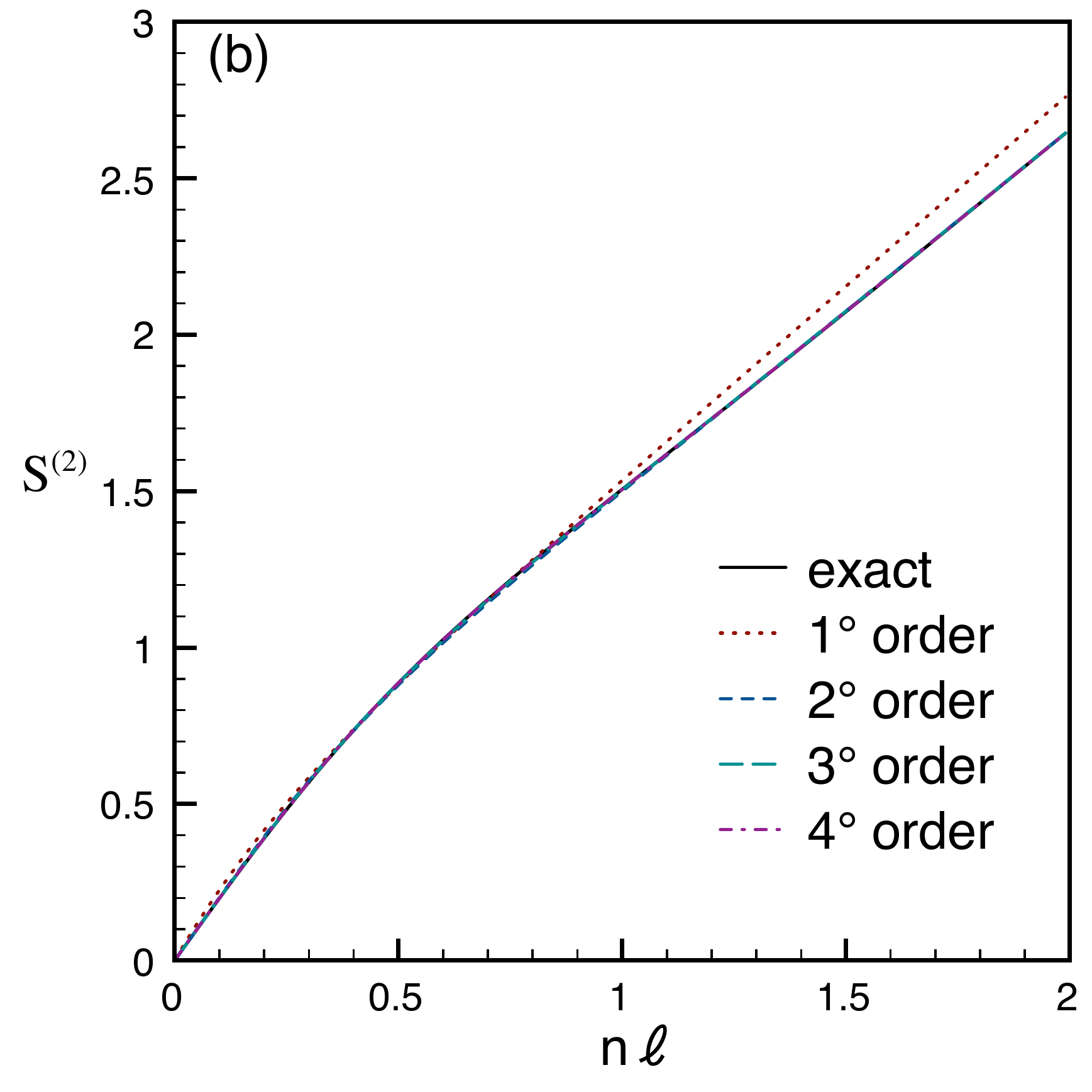}\\
\includegraphics[width=0.48\textwidth]{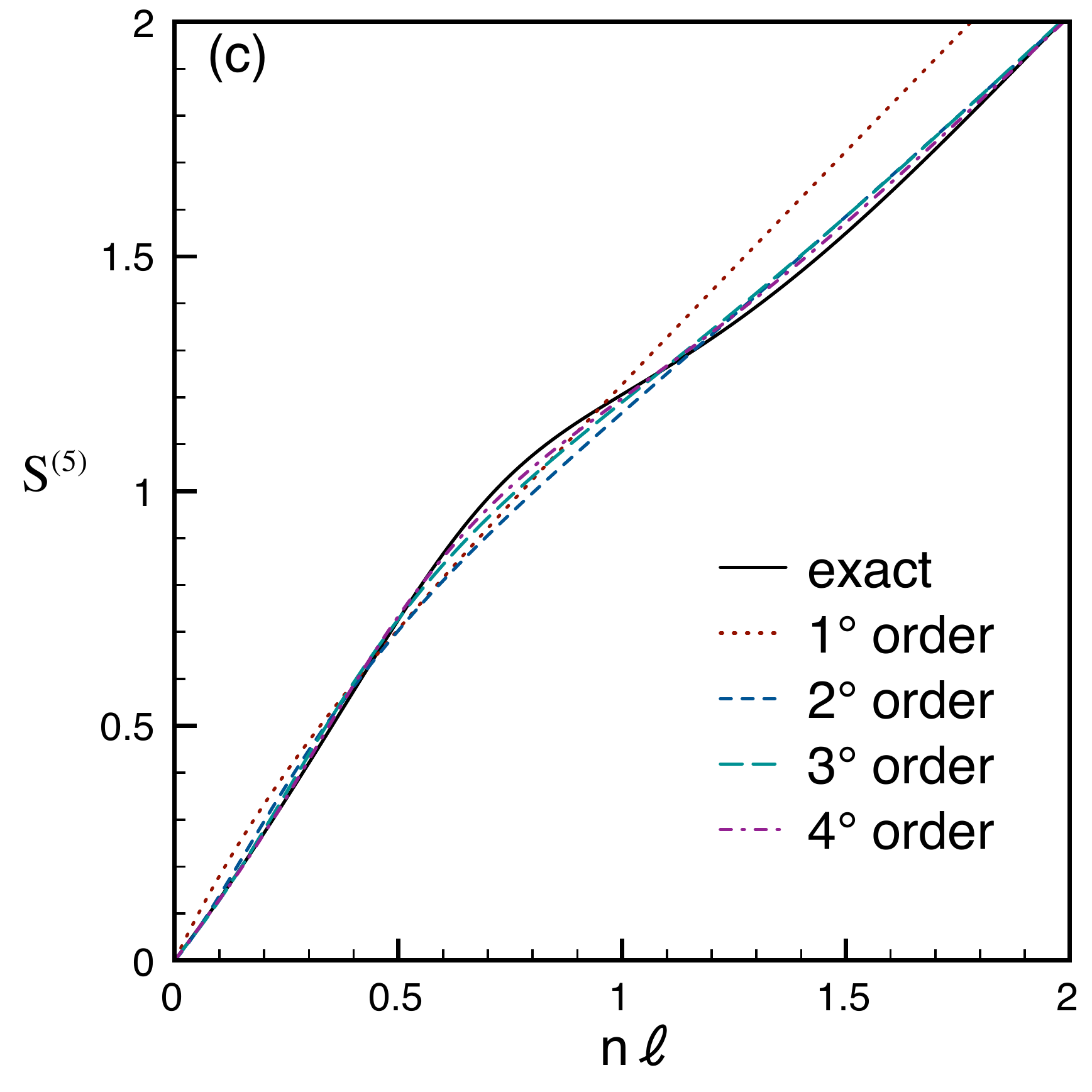}\includegraphics[width=0.48\textwidth]{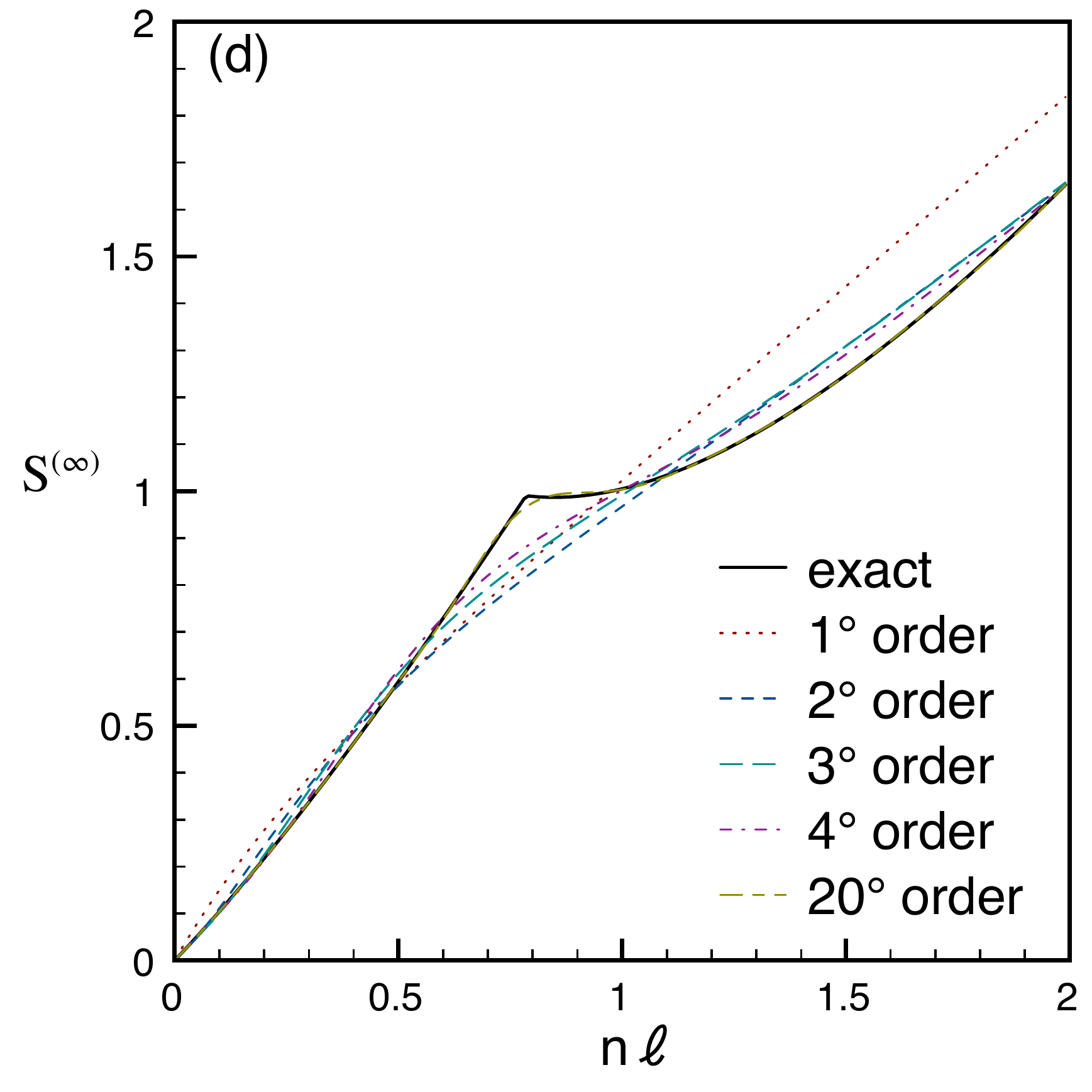}
\caption{Exact entanglement entropy profiles (solid black lines) in the stationary state for a subsystem of length $\ell$ compared with the cumulant expansion given by Eq. (\ref{Salpha_cumulant}) truncated at a given order (dashed lines). While for all integer $\alpha\ge2$ the convergence of the cumulant expansion is very fast, 
for $\alpha=1$ (and all non-integer $\alpha$) the series is only asymptotic and the sum does not converge as it is clear from panel (a).
\label{fig3}}
\end{figure}

Finally, in Fig.\ \ref{fig3} we compare the entanglement entropies reported in Fig. \ref{fig35}, i.e. 
$\alpha=1,2,5,\infty$, with the 
corresponding approximations given by the cumulant expansion (\ref{Salpha_cumulant}) calculated as a sum up to a given 
finite order. 
As an important difference with the ground-state results \cite{song1,song2,cmv-12,v-12b} and some other 
non-equilibrium situations \cite{kl-09,nv-13}, 
all cumulants contribute to the leading behavior of the entanglement entropies and the expansion 
(\ref{Salpha_cumulant}) does not get effectively truncated at the second order. 
As already stressed elsewhere \cite{cmv-11,cmv-12,csc-13}, when all cumulants contribute to the 
expansion (\ref{Salpha_cumulant}), such series is well defined and convergent only for integer $\alpha>1$.
For all other values, and in particular for $\alpha=1$, the coefficients $s^{(\alpha)}_k$ grow 
too quickly with $k$, the resulting series is only asymptotic, and adequate resummation schemes should 
be used to extract quantitative information from it.
In fact, panel (a) in Fig. \ref{fig3} shows that by adding more terms to the expansion (\ref{Salpha_cumulant})
for $\a=1$, we have worse and worse results. 

\subsection{Analytic evaluation of entanglement entropies from the correlation spectrum}

Interestingly, both the leading and subleading behavior of all the traces, cumulants and entropies 
can be analytically extracted by analysing the roots of Eq.\ (\ref{eigenvalues_eqs}) in the limit $n\ell\gg1$. 
For $m\ll n\ell$ one gets $\Omega_{m} \sim \pi m / (2 n\ell)$. However, for any fixed $n\ell$ for large $m$ we have $\Omega_{m} \sim \pi (m-1) / (2 n\ell)$, i.e. there is a $\pi/(2n\ell)$ shift of the roots as $m$ grows. To capture this behavior, we start by solving Eq.~\eqref{equation} perturbatively for large $n\ell$:
\begin{multline}
\Omega_m=\frac{m\pi}{2n\ell}-\frac1{n\ell}\left(\frac{m\pi}{2n\ell}-\frac{m\pi}{2(n\ell)^2}+\frac{\frac{m\pi}2-\frac1{24}(m\pi)^3}{(n\ell)^3}-\frac{\frac{m\pi}2-\frac16(m\pi)^3}{(n\ell)^4}+\right.\\
\left.+\frac{\frac{m\pi}2-\frac5{12}(m\pi)^3+\frac1{160}(m\pi)^5}{(n\ell)^5}+\dots\right)\,.
\end{multline}
The subseries corresponding to the highest powers of $m$ in each term can be summed up:
\be
\Omega_m=\frac{m\pi}{2n\ell}-\frac1{n\ell}\arctan\left(\frac{m\pi}{2n\ell}\right)+
\frac1{(n\ell)^2}\frac{\arctan\left(\frac{m\pi}{2n\ell}\right)}{1+\left(\frac{m\pi}{2n\ell}\right)^2}+\dots\approx\frac{m\pi}{2n\ell}-\frac1{n\ell}\arctan\left(\frac{m\pi}{2n\ell}\right)\,.
\ee
Note that the $\arctan$ function interpolates between $0$ and $\pi/2$, reproducing the expected shift of the solutions
as $m$ grows.
In the large $n\ell$ limit $\Omega$ becomes a continuous variable and all the sums over $m$ can be replaced by integrals. We can compute the density of roots of the equation:
\begin{multline}
\sigma(\Omega_m)=\frac1{\Omega_{m+1}-\Omega_m}=\frac{n\ell}{\frac\pi2+\arctan\left(\frac{m\pi}{2n\ell}\right)-
\arctan\left(\frac{(m+1)\pi}{2n\ell}\right)}\approx\\
\frac{2n\ell}{\pi}\left(1+\frac1{n\ell}\frac{1}{1+\left(\frac{m\pi}{2n\ell}\right)^2}\right)\approx
\frac{2n\ell}{\pi}\left(1+\frac1{n\ell}\frac{1}{1+\Omega_m^2}\right)\,.
\end{multline}
Therefore for $n\ell\gg 1$, using $\lambda_m=(1+\Omega_m^2)^{-1}$ (cf. Eq. (\ref{speclam})), one has
\begin{eqnarray}
\mathrm{Tr}\,\mathbb{C}^k_{A} &=& \int_0^\infty \frac{\ud\Omega}{(1+\Omega^2)^k} \sigma(\Omega )\\ 
&\approx & \frac{2 n\ell}{\pi} \int_{0}^{\infty}\frac{\ud\Omega}{(1+\Omega^2)^k}
\left(1+\frac1{n\ell}\frac{1}{1+\Omega^2}\right)-\frac12
 =  \frac{\Gamma(k-\frac12)}{\sqrt{\pi}\,\Gamma(k)} \left(n\ell+1-\frac1{2k}\right)-\frac12\,,\nonumber
\end{eqnarray}
where the extra $-1/2$ comes from the $m=0$ boundary term which needs to be taken into account when converting the 
sum to integral according to the Euler--Maclaurin formula. 
The above result coincides with the conjectured equation (\ref{conjCk}) providing an explicit proof of it.

For the R\'enyi entropies, using Eq. (\ref{entropies}), we find
\begin{eqnarray}
S^{(\alpha)}_{A} &=& \int_0^\infty \ud\Omega\,\sigma(\Omega ) e_\alpha\Big(\frac1{1+\Omega^2} \Big)
\nonumber\\
&\approx& \frac{2 n\ell}{\pi (1-\alpha)}\int_{0}^{\infty}\ud\Omega\,
 \left(1+\frac1{n\ell}\frac{1}{1+\Omega^2}\right) \nonumber
 \ln\left[\frac{1}{(1+\Omega^2)^{\alpha}}+\left(\frac{\Omega^{2}}{1+\Omega^2}\right)^{\alpha} \right]
 \\&=&  \frac{2 n\ell}{\pi (1-\alpha)}\int_{0}^{\infty}\ud\Omega\,  \ln \frac{1+\Omega^{2\a}}{(1+\Omega^2)^{\alpha}} 
 + \frac{2 }{\pi (1-\alpha)}\int_{0}^{\infty}\ud\Omega\frac{1}{1+\Omega^2}  \ln \frac{1+\Omega^{2\a}}{(1+\Omega^2)^{\alpha}} ,
 \label{Saome}
\end{eqnarray}
and for the von Neumann entropy
\be
\label{S1int}
\begin{split}
S_{A} &\approx \frac{2 n\ell}{\pi}\int_{0}^{\infty}\ud\Omega\,
\left(1+\frac1{n\ell}\frac{1}{1+\Omega^2}\right)
 \left[\frac{1}{(1+\Omega^2)} \ln\left(\frac{1}{1+\Omega^2}\right)+\frac{\Omega^{2}}{1+\Omega^2}
 \ln\left(\frac{\Omega^{2}}{1+\Omega^2}\right) \right]\\
&= 2\,n\ell + (2\ln2-1)\,.
\end{split}
\ee
The single copy entanglement is
\be
\begin{split}
S^{\infty}_{A} &\approx \frac{2 n\ell}{\pi}\int_{0}^{\infty}\ud\Omega\,
\left(1+\frac1{n\ell}\frac{1}{1+\Omega^2}\right)
 \left(\frac12+\left|\frac{1}{1+\Omega^2}-\frac12\right| \right)\\
&= \left(2-\frac4\pi\right)n\ell + 2\ln2-\frac4\pi C\,,
\end{split}
\ee
where $C\approx0.915966$ is Catalan's constant.

Notice that the analytical form as function of $\alpha$ of Eqs. (\ref{Saome})
and (\ref{SAafin}) are apparently very different both for the leading and the subleading term. 
We checked by explicit numerical computation of the two integrals for many different 
(integer and arbitrary real) values of $\alpha$ that they are the same as they should be. 

\section{Comparison with thermodynamic entropies}

In this last section we explore the connection of the entanglement entropy with the thermodynamic one.
At a first look this relation can sound awkward since we are dealing with the time evolution from a pure state,
which is always a pure state and its global entropy should be just zero. 
However, the expectation values of local operators and their correlation functions can be obtained as averages over a proper statistical ensemble which is expected to be thermal for generic systems and the GGE for an integrable one.
For the quench studied in this paper, the convergence to GGE has been established in Ref. \cite{kcc-13}.
Before presenting the calculation of the thermodynamic entropies let us recall how and in which sense 
a thermodynamic ensemble describes the steady state.

For a quench in a general integrable model, the GGE for the whole system is defined as \cite{gg}
\be
\hat \rho_{\text{GGE}}=  \frac{e^{-\sum \lambda_i \hat I_i}}Z\,, 
\label{rGGE}
\ee
where $\{\hat I_i\}$ is a complete set of {\it local} \cite{CEFII,fe-13} integrals of motion, 
$Z= {\rm Tr}\, e^{-\sum \lambda_i \hat I_i}$ is a normalization constant, and the 
Lagrange multipliers $\lambda_i$ are fixed by the initial condition $|\psi_0\rangle$ as  
$\langle \psi_0| \hat I_i |\psi_0\rangle={\rm Tr} [\hat\rho_{\text{GGE}}\hat I_i]$. 
For the quench we are studying in this paper, the final Hamiltonian has a simpler infinite set of conserved charges, 
formed by the fermionic mode occupations, $\hat n(k)$, which are not local. 
Fortunately, the local conserved charges can be expressed as {\it linear} combinations of the 
$\hat n(k)$ \cite{fe-13,csc-13}, so the GGE's built from $\hat n(k)$ and $\{\hat I_i\}$ are equivalent. 

As discussed in Refs. \cite{cdeo-08,bs-08,CEFII,fe-13,f-13} the steady state after a quantum quench in an integrable 
system is described by $\hat\rho_{\text{GGE}}$ in Eq. (\ref{rGGE}) in the sense that in the TDL, a long time limit 
of the reduced density matrix of any finite subsystem $A$ exists and it is equal to the reduced density matrix of $\hat\rho_{\text{GGE}}$.
In formulas, one define (when it exists)
\be
\hat\rho_A^\infty= \lim_{t\to\infty} {\rm Tr}_{B} \big[\lim_{L\to\infty} |\psi(t)\rangle\langle \psi(t)|\big]\,,
\label{rhoAinf}
\ee
and 
 \be
\hat\rho_{A,GGE}= {\rm Tr}_{B} \hat\rho_{\text{GGE}}\,,
\ee
where, in both cases, $B$ is the complement of $A$.
At this point, it is usually said that a system is described by the GGE if $\hat\rho_{A,GGE}=\hat\rho_A^\infty$ \cite{CEFII}.
In Eq. (\ref{rhoAinf}), the order of limits and partial trace is fundamental for the existence of a stationary value. 

Alternatively, the steady state can be described by the so-called diagonal ensemble \cite{p-11}
\be
\hat\rho_\text{d}=\sum_j |c_j|^2|j\rangle\langle j|\,,
\ee
where $c_j=\langle j|\Psi_0\rangle$ are the overlaps of the eigenstates $j$ of the post-quench Hamiltonian with the initial state.
The diagonal ensemble clearly describes the time-averaged values of all observables, including non-local and 
non-stationary ones. In some sense, $\hat\rho_\text{d}$ contains much more information about the quench than the GGE 
which knows only about local observables. 
Indeed it has been argued that in general $\hat\rho_\text{d} \neq \hat \rho_{\text{GGE}}$ \cite{f-13}
while ${\rm Tr}_B \hat\rho_\text{d}= \hat\rho_{A,GGE}$, for any finite $A$.

The inequivalence of the diagonal and GGE ensembles is indeed captured in an easy way by their entropies, 
reflecting the fact that the crucial difference is the information loss in passing from diagonal to GGE ensembles. 
The diagonal and GGE entropies are simply the von Neumann entropies of the corresponding density matrices, 
i.e. 
\begin{align}
S_\text{d}&=-\mathrm{Tr}\hat\rho_\text{d}\ln\hat\rho_\text{d} = -\sum_j |c_j|^2 \ln |c_j|^2\,, \\
S_{\text{GGE}}&= -\mathrm{Tr}\hat\rho_{\text{GGE}}\ln\hat\rho_{\text{GGE}}\,.
\end{align}
%
%

In order to calculate these two thermodynamic entropies, we exploit the fundamental property that in integrable 
models, the summation over states in the expectation values of an observable can be recast
as a functional integral over the Bethe ansatz root densities $\rho(\lambda)$ \cite{ce-13},  as in the 
the Yang--Yang approach to equilibrium thermodynamics \cite{yy}. 
For the stationary values of (some) observables after a quantum quench, it has been shown by Caux and Essler \cite{ce-13}
that, in the thermodynamic limit, only the saddle-point over these root contributes, i.e. 
\be
\lim_{t\to\infty} \langle O(t)\rangle = \lim_{L\to\infty} \langle \Phi_s|O|\Phi_s\rangle\,,
\ee
where $|\Phi_s\rangle$ is the saddle-point state, represented in Bethe ansatz by a proper saddle-point 
Bethe roots density $\rho_s(\lambda)$.


For the quench considered in this paper the saddle point density of roots function $\rho_s(\lambda)$ has already been computed as \cite{kcc-13,nwbc-13}
\be
\rho_\text{s}(\lam)=\frac1{2\pi} \frac1{1+\lam^2/(2n)^2}=\frac1{2\pi}\tilde\rho_\text{s}(\lam)\,.
\ee
In thermodynamic Bethe ansatz, the entropy of a Bethe state 
(defined by the density of particles $\tilde\rho(\lam)$ and holes $\tilde\rho_h(\lam)$ with 
$\tilde\rho_t(\lam)=\tilde\rho(\lam)+\tilde\rho_h(\lam)$)
is given by \cite{yy}
\be
S[\tilde\rho]=L\int\frac{\ud \lam}{2\pi}
\left[\tilde\rho_t(\lam)\ln\tilde\rho_t(\lam)-\tilde\rho(\lam)\ln\tilde\rho(\lam)-\tilde\rho_h(\lam)\ln\tilde\rho_h(\lam)
\right]\,. 
\ee
In the case at hand we have  $\tilde\rho(\lam)=\tilde\rho_\text{s}(\lam)$ and $\tilde\rho_h(\lam)=1-\tilde\rho_\text{s}(\lam)$, so that 
the thermodynamic entropy of the GGE is given by
\be
S_{\text{GGE}}=S[\tilde\rho_\text{s}]=-L\int\frac{\ud \lam}{2\pi}\left[\tilde\rho_\text{s}(\lam)\ln\tilde\rho_\text{s}(\lam)+(1-\tilde\rho_\text{s}(\lam))\ln(1-\tilde\rho_\text{s}(\lam))\right] = 2 n L\,.
\label{S66}
\ee

The diagonal entropy can also be calculated in the thermodynamic limit using the explicit expression for the overlaps 
found in Refs.\ \cite{grd-10,nwbc-13}. Denoting with $c_\rho$ the overlap with a state with root density $\rho(\lam)$,
we have 
\be
S_\text{d}=-\int \mathrm{D}\rho\,\e^{{\cal S}_Q[\rho]}c_\rho^2\ln c^2_\rho = -\int \mathrm{D}\rho\,\e^{{\cal S}_Q[\rho]
+2\ln c_\rho}2\ln c_\rho\,, 
\ee
where ${\cal S}_Q[\rho]$ is the quench action of Ref. \cite{ce-13,mc-12b} whose exponential gives the density of states 
of a root configuration. 
For our case the logarithm of the overlaps is  \cite{nwbc-13} 
\be
-2\ln c_\rho =L\left[\int_0^\infty \frac{\ud \lam}{2\pi} \tilde\rho(\lam) \ln \frac{\lam^2}{4n^2}+n\right]\,. 
\ee
In the TDL, also the diagonal entropy is dominated by the saddle-point 
\be
{\cal S_Q}[\rho_s]=-2\ln c_{\rho_s},
\ee
whose contribution is 
\be
S_\text{d}
= -\int \mathrm{D}\rho\,\e^{{\cal S}_Q[\rho]
+2\ln c_\rho}2\ln c_\rho   \approx-2\ln c_{\rho_\text{s}}= {\cal S_Q}[\rho_s]\,,
\ee
showing in particular that the diagonal entropy equals the saddle-point quench action (as expected). 
Thus we finally have 
\be
S_\text{d}= {\cal S_Q}[\rho_s]=n L+ L\int_0^\infty\frac{\ud\lam}{2\pi} \tilde\rho_\text{s}(\lam) \ln\left(\frac{\lam^2}{4n^2}\right) = n L\,.
\ee

At this point some comments are in order. 
The GGE and diagonal entropies are both extensive as they should be, being thermodynamic entropies. 
They are however different from each other; the GGE entropy has exactly the same density as the entanglement entropy in 
Eq. (\ref{S11}), confirming the expectation that for large time after the quench the entanglement entropy 
{\it does become the thermodynamic entropy}. 
The diagonal entropy is instead exactly {\it half} of the entanglement and GGE entropies.
Once again this reflects the fact that $\hat\rho_{\rm d}$ contains much more information 
than the one needed to describe the expectation values of local observables. 
Furthermore the same ratio of 2 between the GGE and diagonal entropies has been also observed 
in previous studies on the transverse field Ising chain \cite{spr-11,g-13,f-13} and it is natural to wonder 
what the precise physical origin and the value of this ratio is for other quenches in integrable models. 
At a qualitative level, the ratio between diagonal and GGE entropy has been explained by the following 
argument in the Ising chain \cite{g-13,polk-priv}. 
For free fermions, the quench creates excitations in pairs of opposite momenta $k$ and $-k$, 
but in the GGE such correlations are neglected (which in Eq. (\ref{S66}) is encoded in the integral with $\lambda$
going from $-\infty$ to $\infty$). 
Indeed they have no influence on the reduced density matrix of a finite subsystem $A$:  
if a particle with momentum $k$ is in $A$, for long enough time, the $-k$ partner is surely outside of $A$ \cite{g-13}.
Quasi-particle excitations are created in pairs even for the quench considered here \cite{grd-10,cl-11,nwbc-13} (and arguably  
in more general circumstances \cite{cc-05,cc-06}), so the previous qualitative argument still applies. 

\section{Conclusions}

We calculated R\'enyi entanglement entropies in the stationary state after a quench from free to hard-core bosons in one dimension,
exploiting the knowledge of the two-point fermionic correlation function obtained in Ref. \cite{kcc-13} and the 
restoration of Wick's theorem for infinite time. 
R\'enyi entanglement entropies are calculated using two different methods. 
First, following the approach introduced in Refs. \cite{cmv-11,cmv-11b,cmv-12}, we directly sum over 
the powers of the reduced correlation matrix obtaining the integer order R\'enyi entropies. 
In this way, at the end of the calculation we find the  analytical continuation to real $\a$ which provide, among the other things, 
the von Neumann entropy.
The second approach is based on the computation of the spectrum of the reduced correlation matrix, which, in the present case,
is enormously simplified by mapping it into an eigenvalue problem of a second-order differential equation.
Both methods allow us to obtain an explicit and analytic form for the leading and subleading terms in $\ell$ of the R\'enyi entanglement entropies for arbitrary $\alpha$.
In particular, for the von Neumann entropy we find the very simple result
\be
S_A= 2 n\ell +2\ln2-1+\mathcal{O}(e^{-4n\ell})\,.
\ee
The approach to the asymptotic behavior is exponentially fast and so it is already reached for relatively small values of $\ell$.
From the technical point of view it would be very interesting to understand if the problem of the spectrum of the 
reduced correlation matrix can be mapped into a simple differential equation even in other instances, as e.g. recently done 
in Ref. \cite{ep-13}. 

We also compared the von Neumann entanglement entropy with the thermodynamic entropies in both the GGE and the 
diagonal ensembles. 
We found that while entanglement and GGE entropies coincide (as expected), the diagonal entropy is the half of the 
other two. 
This can be easily interpreted as the loss of non-local information passing from the diagonal to the GGE ensemble. 
The same factor of two was previously found also for the transverse field Ising chain \cite{g-13,f-13} and it 
is surely an interesting open problem (numerical initiated in Ref. \cite{spr-11}) 
to understand the relation between diagonal and GGE entropies for more 
complicated quantum quenches in integrable models.

\section*{Acknowledgments}   
The authors acknowledge the ERC  for financial  support under  Starting Grant 279391 EDEQS. 
We thank Marcos Rigol and Jacopo De Nardis for very fruitful discussions. 
We are grateful to Ingo Peschel and Anatoli Polkovnikov for correspondence.


\end{document}